\documentclass[12pt,a4paper]{article}

\begin{document}

\title{Orthogonal polynomials, special functions\\and mathematical physics}
\author{Miguel Lorente \\ \small Departamento de F\'{\i}sica, Universidad de Oviedo, 33007 Oviedo, Spain}
\date{ }
\maketitle
%\begin{abstract}
In the 6th Symposium of OPSA, there have been several communications dealing with concrete applications of
orthogonal polynomials to experimental and theoretical physics, chemistry, biology and statistics.
Here, I make suggestions concerning the use of the powerful apparatus of orthogonal polynomials and
special functions in several lines of research in mathematical physics.
%\end{abstract}

\ 

(1) {\em Hilbert spaces}: The postulates of quantum mechanics introduced an abstract space where the states
of a system were represented by function-valued vectors and the physical magnitudes by operators. The
formalism best suited to this model was the recently invented Hilbert space, where the basis was given
by orthogonal polynomials [3]. The newly discovered orthogonal polynomials in the 20th century and the
recently proposed multiple orthogonal polynomials open the door to new applications to more elaborate
quantum models [14,16].

\ 

(2) {\em Integrable models}: A classical or quantum system can be described rigorously and exactly by some
linear or nonlinear differential equation whose solution, except in a few cases, is not known. Very
hard work has been done recently to integrate such a system with the help of inverse scattering
methods, or soliton solutions and lattice approximations [10]. In some cases the solutions of the
non-linear equation can be expressed in terms of solutions of classical special functions (Airy,
Whittaker, Bessel, parabolic cylinder functions [5,6]). In the case of linear equations knowledge of
the newly discovered orthogonal polynomials and the corresponding differential equation can be used to
compare with the equation of the physical system in order to solve it.

\ 

(3) {\em Group representations and special functions}: The symmetries of some physical systems provide a
powerful tool for calculating solutions [9]. To each symmetry there corresponds a group. The carrier
space of some irreducible representation of a group can be expressed with the help of special
functions [15]. In this symposium there have been proposed some particular groups associated to the
Lagrangian, with the help of root systems of a classical Lie group the solution of which are given by
the associated orthogonal polynomials. The class of quantum integrable systems associated with root
systems was introduced by Olshanetsky and Perelomov as a generalization of the Calogero- Sutherland
systems [12]. For the potential $v(q) = k(k - 1)/(sin q)^2$ the wave functions of such a systems are
related to polynomials in $l$ variables (where $l$ is the rank of the root systems) and they are a
generalization of the Gegenbauer and Jack polynomials, that are connected to physical systems of l
particles [13]. In this symposium similar arguments have been proposed for the Weyl group of type $B_n$;
there are associated a Lagrangian and the spherical harmonics proposed by Dunkl. For the root system
$C_n$ there are associated the Macdonald-Koornwinder polynomials. These new methods open the way to
constructing orthogonal polynomials of several variables and the corresponding recurrence relations
and raising and lowering operators.

\ 

(4) {\em Orthogonal polynomials of a discrete variable}: Although the Kravchuk, Meixner, Charlier and Hahn
polynomials of a discrete variable were introduced many years ago [8], the application to physical
models on the lattice has started recently. First, the newly developed formalism of quantum groups,
q-analysis, q-orthogonal polynomials and special functions can be considered as special cases of
orthogonal polynomials on non-homogeneous lattices, and they are widely used in perturbative methods
of quantum field theories [2]. Also the classical orthogonal polynomials and function of discrete
variable have been applied to quantum systems such as the harmonic oscillator, the hydrogen atom and
the Dirac equation on the lattice [4]. For some arbitrary one-dimensional systems, such as the Toda
lattice, the evolution of the system in discrete time can be obtained exactly with the help of the
orthogonal polynomial the argument of which are the symmetric product of the creation and annihilation
operator of the harmonic oscillator. New insights into the solution of Dirac equation on the lattice
have been obtained with the Wilson operators that satisfy a mild condition of chiral invariance [7].
Finally, in quantum gravity, a new area of research, the method of Ponzano-Regge calculus and the
equivalent method of Penrose [11] spin-networks has been worked out with the help of the $3j - n$
symbols, that are connected with the Hahn polynomials of a discrete variable, and are suited for the
discretization of space and time of General Relativity [1].

\end{document}